# Conformal invariance in the three dimensional (3D) Ising model and quaternionic geometric phase in quaternionic Hilbert space

Z.D. Zhang[a] and N.H. March[b,c]

[a] Shenyang National Laboratory for Materials Science, Institute of Metal Research and International Centre for Materials Physics, Chinese Academy of Sciences, 72 Wenhua Road, Shenyang, 110016, P.R. China

[b] Department of Physics, University of Antwerp, Antwerp, Belgium

[c] Oxford University, Oxford, England

Based on the quaternionic approach developed by one of us [Z.D. Zhang, Phil. Mag. 87 (2007) 5309.] for the three-dimensional (3D) Ising model, we study in this work conformal invariance in three dimensions. We develop a procedure for treating the 3D conformal field theory. The 2D conformal field theory is extended to be appropriate for three dimensions, within the framework of quaternionic coordinates with complex weights. The Virasoro algebra still works, but for each complex plane of quaternionic coordinates. The quaternionic geometric phases appear in quaternionic Hilbert space as a result of diagonalization procedure which involves the smoothing of knots/crossings in the 3D many-body interacting spin Ising system. Possibility for application of conformal invariance in three dimensions on studying the behaviour of the world volume of the brane, or the world sheet of the string in 3D or (3+1)D, is briefly discussed.

The corresponding author: Z.D. Zhang, Tel: 86-24-23971859, Fax: 86-24-23891320, e-mail address: zdzhang@imr.ac.cn

It is well known that in the context of a physical system with local interactions, conformal invariance is an immediate extension of scale invariance, a symmetry under dilations of space [1]. Conformal transformations are dilations by a scaling factor that is a function of position (local dilations) [2]. The study of conformal invariance in two dimensions was initiated by Belavin, Polyakov and Zamolodchikov, which combined the representation theory of the Virasoro algebra with the idea of an algebra of local operators [3]. Conformal field theories not only provide toy models for genuinely interacting quantum field theory, but also play a central role in string theory [4] where two-dimensional (2D) scale invariance appears naturally. The vibrations of a string with internal degrees of freedom can be studied within a conformal field theory, from the point view of the world sheet. A classification of the 2D conformal field would provide useful information on the classical solution space of string theory [5]. Conformal field theories also describe critical phenomena at the critical point of the second-order phase transitions in two dimensions, where the correlation length diverges [5]. The partition function of a statistical mechanics model can be closely related to the knot polynomials by matrix elements of the braiding matrices of an associated rational conformal field theory, or alternatively, the matrix elements of the R-matrix of a quantum group [6,7].

The critical point of the 2D Ising model, as a canonical example, is described by a conformal field theory, since every scale-invariant 2D local quantum field theory is actually conformally invariant. The high-temperature disordered phase and the low-temperature ordered phase in the 2D Ising model are related by a duality of the

model, and the second order phase transition occurs at the self-dual point [8]

A finite number of parameters ((d+1)(d+2)/2) are needed to specify a conformal transformation in d spatial dimensions [1]. The consequence of this finiteness is that in three or more dimensions, conformal invariance does not turn out to give much more information than ordinary scale invariance [5]. It is commonly accepted that the exception is in two dimensions, where the number of parameters specifying local conformal transformations is infinite. In two dimensions, the conformal algebra becomes infinite dimensional, leading to significant restrictions on the 2D conformally invariant theories [5]. It means that an infinite variety of conformal transformations exist in 2D, which, although not everywhere well defined, are locally conformal [1]: they are holomorphic mappings from the complex plane (or part of it) onto itself. A local field theory should be sensitive to local symmetries, even if the related transformations are not globally defined. It is local conformal invariance that enables exact solutions of 2D conformal field theories [1]. This is the reason for the success of conformal invariance in the study of 2D critical systems.

One of us (ZDZ) [9] has worked on the three-dimensional (3D) Ising model, using a quaternionic approach, which has recently found favour with mathematicians [10-12]. It was pointed out in [9,13,14] that the framework of the statistical mechanics for 3D Ising magnets should include the time, being in the (3+1) dimensional Euclidean spacetime. This argument is based on a fact that the temperature in statistical mechanics is actually the time in quantum field theory [15]. This is because the Euclidean time interval can be consistently identified with $\beta$. There are serious challenges to the validation of the ergodic hypothesis in the 3D many-body interacting spin systems, like the 3D Ising model, where the topologic contributions to the partition function as well as correlation functions and other physical quantities cannot be negligible [9,13,14]. In a more recent work [16], we represented a detailed analysis of temperature-time duality in the 3D Ising model. It was pointed out that the time necessary for the time averaging must be infinite, being comparable with or even much far than the time of measurement of the physical quantity of interests, because the topologic effects are non-local so that an efficient sweep of the ensemble by any of its microstates should be infinite since the number of these microstates is infinite [16]. Therefore, the integrand of the partition function of the 3D Ising model should be performed in four dimensions, since one needs to take the time average by the integrand in the fourth dimension.

In this work, we develop a procedure for treating the 3D conformal field theory. We study the conformal invariance in the 3D Ising model by the quaternionic approach developed in [9] and discuss the quaternionic geometric phase in quaternionic Hilbert space. We uncover that for treating the 3D conformal field theory, the representation theory of the Virasoro algebra and the algebra of local operators should be combined with the quaternionic geometric phase in quaternionic Hilbert space. The decomposition of 3D conformal blocks to 2D conformal blocks can be done by the utilization of Jordan-von Neumann-Wigner procedure [17].

The success of studying the 3D Ising model within the (3+1)D framework provides a chance to understand deeper conformal invariance and conformal transformations in three dimensions. From another point of view, it is anticipated that the procedure for the exact solution of the 3D Ising model enables local conformal invariance in three dimensions. It is possible to remove the finiteness parameters needed to specify a conformal transformation in three dimensions, so that the conformal algebra becomes infinite dimensional. This implies that, in order to maintain local conformal invariance in three dimensions, the exact solution of the 3D Ising model should possess the character of the exact solution of the 2D Ising model, as what Zhang proposed in [9]. Indeed, the exact solution of the 3D Ising model, followed by means of two conjectures, has the main feature of the Onsager's exact solution of the 2D Ising model, while the only difference between the solutions in 3D and 2D is the appearance of weight factors in the partition function and eigenvectors within the (3+1)D quaternionic framework [9].

The conformal group in d-dimension is the subgroup of coordinate transformations that leaves the metric invariant up to a scale change [1,3-5],

$$g_{\mu\nu}(x) \to g'_{\mu\nu}(x') = \Lambda(x) g_{\mu\nu}(x) \tag{1}$$

Here we consider the space $R^d$ with flat metric $g_{\mu\nu} = \eta_{\mu\nu}$ with signature (p,q) and line element $ds^2 = g_{\mu\nu} dx^\mu dx^\nu$. For d > 2, the conformal algebra is isomorphic to SO (p+1, q+1) [5]. Actually, the conformal group admits a nice realization acting on $R^{p,q}$, stereographically projected to $S^{p,q}$, and embedded in the light cone of $R^{p+1,q+1}$ [5].

The 2D conformal field theory is naturally defined on a Riemann surface (or complex curve), i.e., on a surface possessing suitable complex coordinates. In the case of the sphere, the complex coordinates can be taken to be those of the complex plane that cover the sphere except for the point at infinity. The quaternionic approach developed in [9] provides a possibility to deal with the 3D conformal field theory in the bases of the Hilbert space with suitable quaternionic coordinates. Here we suggest that the Hilbert space of the 3D Ising model, so-called complexified quaternionic Hilbert space, provides the important information of the conformal field theory in three or (3+1) dimensions. So the correlation functions in the 3D Ising systems and also the 3D conformal field theory depend on the quaternionic parameters. The conformal invariance in the 3D Ising model can be studied also with the quaternionic coordinates. In what follows, we will illustrate how to process the conformal transformation in three dimensions.

For a 3D many-body interacting system, like the 3D Ising model or the 3D conformal field theory, we need to decompose 3D conformal blocks to 2D conformal blocks, by the utilization of Jordan-von Neumann-Wigner procedure [17], which can be realized by introducing an additional dimension, performing a unitary transformation as a rotation in higher dimensions, and constructing a quaternionic basis together with weight factors [9]. The normalized eigenvectors (Eq. (33) in [9]), proposed for the 3D Ising model, are quaternionic eigenvectors (also see [10-12] for the mathematical outlook). The complexified weight factors have significance of topological phases as revealed in [9,13,14]. Actually, the quaternionic bases found for the 3D Ising model in [9] are complexified quaternionic bases. One refers to [18] for details of complexified quaternion, [19-21] for quaternionic quantum mechanics, and [22] for quaternion and special relativity. The use of Clifford structures and the P.

Jordan structures can make the way of applying the quaternion structure more elegant and simpler [10-12]. Jordan algebras with its multiplication $A \circ B = \frac{1}{2}(AB + BA)$ instead of the usual matrix multiplication AB replaces in an elegant way the desire of looking for commutative subalgebras of the algebra constructed and for combinatorial properties. Such desire is one of the main obstacles in solving exactly the 3D Ising model. We believe that it is also the case for the 3D conformal field theory.

For the 3D conformal field theory, therefore, one can define the quaternionic coordinates $q = x^0 + iw_1 x^1 + jw_2 x^2 + kw_3 x^3$ and $\bar{q} = x^0 - iw_1 x^1 - jw_2 x^2 - kw_3 x^3$. Here $w_i$ (i = 1, 2, 3) are complex weight factors on the imaginary coordinates of the quaternionic bases. The 3D conformal transformations can thus be decomposed into three 2D conformal transformations that coincide with the analytic coordinate transformations $z_i \to f_i(z_i)$ and $\bar{z}_i \to f_i(\bar{z}_i)$ (i = 1, 2, 3), the local algebra of each of which is infinite dimensional as what we have for the 2D conformal field theory. In the complexified quaternionic coordinates, we have

$$ds^2 = dq d\bar{q} \to \sum_{i=1}^{3} \left|\frac{\partial f_i}{\partial z_i}\right|^2 |w_i|^2 dz_i d\bar{z}_i \qquad (2)$$

Therefore, the 2D conformal field theory can be generalized to the 3D conformal field theory in such a way: Each imaginary coordinate in the quaternionic coordinates and the real coordinate construct the complex coordinates for the 2D conformal field theory. We can study the conformal invariance within such complex coordinates, keeping in mind that there are complex weight factors $w_i$ (i = 1, 2, 3) for each imaginary coordinate in the quaternionic coordinates. The differences between eq. (2)

for the 3D conformal field theory and that for the 2D conformal field theory in literature [1,3-5] are as follows: 1) the summation i and 2) the phase factors $w_i$.

The conformal transformation in three dimensions can be written as

$$\tau_i \mapsto A_i \tau_i = \frac{a_i \tau_i + b_i}{c_i \tau_i + d_i} \tag{3}$$

with the matrix

$$A_i = \begin{pmatrix} a_i & b_i \\ c_i & d_i \end{pmatrix} \quad \text{where} \quad a_i d_i - b_i c_i = 1,$$

Again, i = 1, 2 or 3 corresponds to the complex plane constructed by the real coordinate and one of the three imaginary coordinates of quaternionic bases. Although each conformal transformation in eq. (3) for three dimensions has the same feature as the conformal transformation for two dimensions in literature [1,3-5], one should notice that there are three different conformal transformations performed in the quaternionic bases for a 3D system.

Following the procedure for the 2D conformal field theory [1,3-5], the energy-momentum tensor in the 3D systems (in (3+1)D framework) can be expanded as

$$L(z) = \sum_{i=1}^{3} \sum_{n=-\infty}^{\infty} |w_i| \operatorname{Re} |e^{i\phi_i}| L_{ni} z_i^{-n-2} \tag{4}$$

with the commutator as

$$[L_{m_i}, L_{n_i}] = (m_i - n_i) L_{m_i + n_i} + \frac{c}{12} m_i (m_i^2 - 1) \delta_{m_i + n_i, 0} \tag{5}$$

here i = 1, 2, 3. Once again, the summation i and the phase factors $w_i$ appear in eq. (4) for the 3D conformal field theory, which do not exist in that for the 2D conformal field theory in literature [1,3-5]. In eq. (4), however, because only the real part of the phase factors appears, the complex weight factors $w_i$ are replaced by $|w_i| |\text{Re}|e^{i\phi_i}||$, where $\phi_i$ are phases [see [13] for details of topological phases in the 3D Ising model]. The famous Virasoro algebra still works, but here for each complex plane of quaternionic coordinates in the complexified quaternionic Hilbert space. The parameter c is the central charge, which is the same for different complex planes of a model, c = 1/2 for the Ising model. Note that if the theory contains a Virasoro field, the states transform in representation of the Virasoro algebra, rather than just the Lie algebra of sl(2, C) corresponding to the Möbius transformation [4].

For the 2D conformal field theory, the scaling dimension is given by $\Delta = h + \bar{h}$, here h and $\bar{h}$ are known as the conformal weights of the state, which are independent (real) quantities. We have $h = \bar{h} = \frac{1}{16}$, and $\Delta = \frac{1}{8}$, and there is a relation between the scaling dimension $\Delta$ and the critical exponent $\eta$, based on the formalism of the correlation function: $\eta = 2\Delta + 2 - d$. For the 2D Ising model, $\eta$ = 1/4 and $\Delta$ = 1/8. For the 3D Ising model, the critical exponent $\eta$ is found to be 1/8 [9]. Thus, one has $\Delta$ = 9/16.

Then, we discuss the physical essential of the phase factors $w_i$ appearing in eq. (2) and eq. (4) above for the 3D conformal transformation. In the physics of gauge theories, Wilson lines correspond essentially to the space-time trajectory of a charged particle, i.e., so-called world histories of mesons or baryons [6]. Under a change of framing, the expectation values of Wilson lines are multiplied by a phase exp(2$\pi$i$h_a$), where $h_a$ is the conformal weight of the field. A twist of a Wilson line is equivalent to a phase, while a braiding of two Wilson lines from a trivalent vertex is also equivalent to a phase. The skein relation for Wilson lines in the defining N dimensional representation of SU(N) can be found in [6]. In [9,13,14], Zhang proposed that the complex weight factors, i.e., the topological phase factors exist in the 3D Ising model. These topological phase factors may have the same origin of the phase in the expectation values of Wilson lines obtained by a change of framing in gauge theories. This is because the procedure for diagonalization involves the smoothing of knots/crossings in the 3D Ising system. Therefore, the phase factors $w_i$ appearing in eq. (2) and eq. (4) above for the 3D conformal transformation may have the same origin as those in the expectation values of Wilson lines [as well as the 3D Ising model]. For further understanding, one may refer also to [23] for a nonadiabatic geometric phase in quaternionic Hilbert space, and [24] for a quaternionic phase and coherent states in quaternionic quantum mechanics.

In 2D conformal field theory, canonical quantization on a circle S gives a physical Hilbert space $H_s$. A vector $\psi \in H_s$ is a suitable functional of appropriate fields on S, which corresponds to a local field operator $O_\psi$. There is a relation in conformal field theory between vectors in Hilbert space and local operators. A 3D analog of such relation between states and local operators can be found also, as shown in [6]. However, according to the quaternionic approach developed for the 3D Ising model, some new features are uncovered for relation between states and local operators in 3D systems. Usually, the physical Hilbert space obtained by quantization in (2+1) dimensions can be interpreted as the spaces of conformal blocks in (1+1) dimensions [25] Analogously, the physical Hilbert space obtained by quantization in (3+1) dimensions can be interpreted still as the spaces of conformal blocks in three (1+1) dimensional complex planes of the quaternion coordinates.

We summarize here the procedure we developed above for treating the 3D conformal field theory: 1) Introducing an additional dimension to construct a (3+1)-dimensional framework to form the quaternionic coordinates; 2) Performing a unitary transformation, as a rotation in (3+1)-dimensions, to represent states and operators in the (3+1)-dimensional complexified quaternionic Hilbert space; 3) while introducing complex weight factors as topological phase factors $w_i$, for smoothing knots/crossings; 4) the decomposition of 3D conformal blocks to 2D conformal ones; 5) Dealing with 2D conformal blocks in each complex plane of quaternionic coordinates in the complexified quaternionic Hilbert space; 6) accounting the summation i of 2D conformal blocks in three complex planes together with the contributions of the phase factors $w_i$.

Next, we discuss briefly the connection with the string theory: The world sheet is the 2D surface that the string sweeps out as it propagates through space-time, which can be described by a 2D conformal field theory. Namely, conformal theories on the plane are often considered as string vacua, the nonfluctuating flat space version of some string theories, serving as toy models for quantum string theory [1]. The square Ising model on a closed Riemann surface with deformed squares, representing positive or negative curvature, is instrumental in investigations of quantum gravity of the coupling of a matter (Ising) theory to fluctuations of space-time geometry An interesting problem is how to understand the behaviour of the brane as it sweeps out in space-time. Our understanding is: we need to describe of the world volume of the brane in the (3+1) D framework as what one of us (ZDZ) did in [9] for the 3D Ising model. In this sense, suitable quaternionic coordinates with complexified weight factors should be introduced and then conformal invariance in three dimensions (actually in (3+1) D) can be utilized to study the behaviour of the world volume of the brane. This can be related with the observable universe, which could be a brane, i.e., a (3+1) surface, embedded in the bulk, i.e, a (3+1+d) – dimensional spacetime, with standard – model particles and fields trapped on the brane, while gravity is free to access the bulk [26].

In conclusion, quaternion-based functions developed in [9] for the 3D Ising models, which is related to the quaternionic sequence of Jordan algebras implied by the fundamental paper of Jordan, von Neumann, and Wigner [17], can be utilized to study the conformal invariance in dimensions higher than two. The 2D conformal field theory can be extended to be appropriate for three dimensions, within the framework of the quaternionic coordinates with complex weights. The 3D conformal transformations can be decomposed into three 2D conformal transformations, where the Virasoro algebra still works, but only for each complex plane of quaternionic coordinates in the complexified quaternionic Hilbert space. Then one needs to perform the summation i of 2D conformal blocks in three complex planes together with the contributions of the phase factors $w_i$. Similar to 2D conformal field theories, local conformal invariance in 3D (though it is limited in each complex plane of quaternionic coordinates) enables exact solutions of 3D conformal field theories. The success of studying 3D critical systems, such as the 3D Ising model in [9], is due to this conformal invariance. The scaling dimension $\Delta$ is predicted to be 9/16 for the 3D Ising model, to be contrasted with the known value 1/8 for the 2D Ising model. We discuss the physical essential of the phase factors $w_i$ appearing in eq. (2) and eq. (4) for the 3D conformal transformation, and suggest that they may have the same physical significance with those in the expectation values of Wilson lines [as well as the 3D Ising model]. Possibility of utilizing the present results for conformal invariance in three dimensions and quaternionic geometric phase in quaternionic Hilbert space for studying the behaviour of the world volume of the brane, or the world sheet of the string in 3D or (3+1)D, is briefly discussed. The present work is helpful for better understanding classical, conformal and topological field theories in high dimensions [27,28].

**Acknowledgements**

ZDZ acknowledges the support of the National Natural Science Foundation of China under grant number 50831006. NHM acknowledges on-going affiliation with the University of Antwerp (UA) plus partial financial support through BOF-NOI, made possible by Professors D. Lamoen, and C. Van Alsenoy.

Naber and T.S. Tsun, (Elsevier Inc. 2007).